\documentclass[12pt]{article}
\begin{document}
\title{A Remark on BRST Singlets}
\author{E. Kazes
~\\~\\~\\
Department of Physics\\
104 Davey Laboratory\\
University Park, PA 16802}
\maketitle
\begin{abstract}
Negative norm Hilbert space state vectors can be BRST invariant, we show in a simplified Y-M model that such states can be created by starting with gluons only. 
\end{abstract}

\newpage
\addtolength{\baselineskip}{\baselineskip}

Faddeev and Slavnov\cite {fs} pointed out that BRST\cite{brs,tl} invariance alone is not sufficient for a Y-M type theory to be physically admissible since ``physical" states that are annihilated by the generators of BRST transformation may have negative norm.  They emphasized the need for additional investigation of the BRST-singlet sector.  By using a second quantized F-P\cite{fp} model I find that the transition amplitude from an initial BRST singlet of positive norm to a final state of negative norm is different from zero.  There is no barrier to the production of states of negative norm.

The minimal Y-M Lagrangian for gluons, ghosts and antighost fields is
\begin{equation}
{\cal L} = {\cal L}_{o}(A)+{\cal L}_{GF}+{\cal L}_{FP}\\
\end{equation}
\begin{equation}
{\cal L}_{o}=-\frac{1}{4}F^{a}_{\mu\nu}F^{\mu \nu a}\\
\end{equation}
\begin{equation}
{\cal L}_{GF}=-\frac{1}{2}(\partial_{\mu}A^{\mu a})^2\\
\end{equation}
\begin{equation}
{\cal L}_{FP} = - i \partial^{\mu} \overline{C}^{a} (D_{\mu}C)^a\\
\end{equation}
\begin{equation}
D_{\mu} = \partial_{\mu} - i g A_{\mu}\\
\end{equation}

The lowest order transition amplitude from an initial state with two gluons to a final ghost, anti-ghost state requires two types of Feynman diagrams. The one, with two gluon-ghost vertices gives the amplitude
\begin{equation}- \frac{i}{2}g^{2} \left[p_{1} \cdot \varepsilon (k_{1}) p_{2} \cdot \varepsilon(k_{2})  C_{a_{1}b_{1}f}  C_{a_{2}b_{2}f} + p_{1} \cdot \varepsilon (k_{2}) p_{2} \cdot \varepsilon(k_{1})   C_{a_{2}b_{1}f}  C_{a_{1}b_{2}f} \right]\\
\end{equation}

The other, which has one gluon-ghost and a three gluon vertex, gives
\begin{equation}\frac{ig^{2}}{2k_{1}\cdot k_{2}} \left[\varepsilon(k_{1}) \cdot \varepsilon (k_{2}) (k_{1}-k_{2}) \cdot p_{1} - 2 p_{1} \cdot \varepsilon (k_{1}) k_{1} \cdot \varepsilon(k_{2}) + 2 p_{1} \cdot \varepsilon (k_{2}) k_{2} \cdot \varepsilon (k_{1}) \right] C_{a_{1}a_{2}f} C_{b_{1}b_{2}f}\\
\end{equation}

where $C_{abc}$ are the structure constants of the group $SU(n)$. Repeated indices are summed.  The incident gluon momenta, polarization and color indices respectively are\\
\noindent
$k_{1},\varepsilon(k_{1}),a_{1}; k_{2}, \varepsilon(k_{2}),a_{2}$.  The ghost, antighost momenta, and color indices are $p_{1}, b_{1};p_{2},b_{2}$.  A common feature of 
ghost production amplitudes is their gauge dependence, this will be seen later to follow directly from BRST invariance.  The replacement $\varepsilon(k)\rightarrow\varepsilon(k)+\lambda k$ changes the sum of (6) and (7), whereas ghost free amplitudes are invariant under this residual gauge transformation.

From the anticommutation relations of ghost and antighost fields it follows that the final state in this process
\begin{equation}
\Psi=\overline{c}_{b_{1}}^{+}(p_{1})c_{b_{2}}^{+}(p_{2})\mid  0 \rangle\\
\end{equation}
has zero norm if $p_{1}\neq p_{2}$.  Whereas for smeared ghost, antighost states
\begin{equation}
\Psi^{'}=\int \underline{d}p_{1} \underline{d}p_{2} f(p_{1}) g(p_{2}) \overline{c}_{b_{1}}^{+}(p_{1})c_{b_{2}}^{+}(p_{2})\mid 0 \rangle\\
\end{equation} 
the norm is
\begin{equation}
\langle \Psi^{'} \mid \Psi^{'} \rangle = - \mid \int f(p_{1}) g(p_{2}) \underline{d}p_{1} \underline{d}p_{2} \mid^{2} \delta_{b_{1}b_{2}}\\
\end{equation}
A negative norm results when $b_{1}=b_{2}$ for an appreciable overlap of the two momentum distribution $f$ and $g$.  Therefore the sum of (6) and (7), with $p_{1}\neq p_{2}$, is the amplitudes of a zero norm vector in Fock space.

The amplitude in Eq.(7) is ambiguous for $p_{1} \rightarrow p_{2}$ since then $k_{1} \rightarrow k_{2} \rightarrow p_{2}$.  We avoid this by including a gluon in the final state with momenta, polarization and color index $k_{3}, \varepsilon (k_{3}), a_{3}$ respectively.  To obtain all amplitudes of order $g^{3}$ requires the quartic and cubic gluon-gluon vertices as well as a gluon-ghost interaction.  For $b_{1}=b_{2}=b$ the amplitude which involves the quartic vertex vanishes.  For $p_{1}=p_{2}=p$ the diagram that has three gluon-ghost vertices gives
\begin{displaymath}
\frac{g^{3}}{4} \left[ \frac{1}{p \cdot k_{1} \, \, p \cdot k_{3}} p \cdot \varepsilon (k_{1}) p \cdot \varepsilon (k_{3})  (k_{1} + k_{3}) \cdot \varepsilon (k_{2})  C_{a_{1}bf} C_{a_{2}fg} C_{a_{3}bg}
\right .
\end{displaymath}
\begin{displaymath}
+ \frac{1}{p \cdot k_{2} \, \, p \cdot k_{3}} p \cdot \varepsilon (k_{2}) p \cdot \varepsilon (k_{3})  (k_{2} + k_{3}) \cdot \varepsilon (k_{1})  C_{a_{2}bf} C_{a_{1}fg} C_{a_{3}bg}
\end{displaymath}
\begin{equation}
\left .
- \frac{1}{p \cdot k_{1} \, \, p \cdot k_{2}} p \cdot \varepsilon (k_{1}) p \cdot \varepsilon (k_{2})  (k_{1} - k_{2}) \cdot \varepsilon (k_{3})  C_{a_{1}bf} C_{a_{3}fg} C_{a_{2}bg} \right]
\end{equation}

where all repeated indices except $b$ are summed.  In order to show that the product of the three structure constant in Eq.(11) does not vanish identically it is convenient to sum over $b$, and we obtain $C_{a_{1}a_{2}a_{3}}$.

For the same process, the amplitude that contains a gluon-ghost and the cubic gluon interaction gives
\begin{displaymath}
\frac{g^{3}}{2} \, \frac{p \cdot \varepsilon(k_{1})}{p \cdot k_{1} \,\, k_{2} \cdot k_{3}} \left[ \varepsilon(k_{2}) \cdot \varepsilon(k_{3}) \, (k_{2} + k_{3}) \cdot p - 2 p \cdot \varepsilon(k_{3}) k_{3} \cdot \varepsilon (k_{2})
\right .
\end{displaymath}
\begin{displaymath}
\left .
- 2p \cdot \varepsilon(k_{2}) k_{2} \cdot \varepsilon(k_{3}) \right] C_{a_{1}br} C_{rbl} C_{la_{2}a_{3}}
\end{displaymath}
\begin{displaymath}
+ \frac{g^{3}}{2} \, \frac{p \cdot \varepsilon(k_{2})}{p \cdot k_{2} \,\, k_{1} \cdot k_{3}} \left[ \varepsilon(k_{1}) \cdot \varepsilon(k_{3}) \, (k_{1} + k_{3}) \cdot p - 2 p \cdot \varepsilon(k_{3}) k_{3} \cdot \varepsilon (k_{1})
\right .
\end{displaymath}
\begin{displaymath}
\left .
- 2p \cdot \varepsilon(k_{1}) k_{1} \cdot \varepsilon(k_{3}) \right] C_{a_{2}br} C_{rbl} C_{la_{1}a_{3}}
\end{displaymath}
\begin{displaymath}
+ \frac{g^{3}}{2} \, \frac{p \cdot \varepsilon(k_{3})}{p \cdot k_{3} \,\, k_{2} \cdot k_{1}} \left[ \varepsilon(k_{2}) \cdot \varepsilon(k_{1}) \, (k_{2} - k_{1}) \cdot p + 2 p \cdot \varepsilon(k_{1}) k_{1} \cdot \varepsilon (k_{2})
\right .
\end{displaymath}
\begin{equation}
\left .
- 2p \cdot \varepsilon(k_{2}) k_{2} \cdot \varepsilon(k_{1}) \right] C_{a_{3}br} C_{rbl} C_{la_{2}a_{1}},
\end{equation}
and since
\begin{equation}
\sum_{b,r} \, C_{mbr} \, C_{mbr} \, \, \alpha \, \delta_{mn}
\end{equation}
it follows that the products of the structure constants is not zero for every $a_{1}, a_{2}, a_{3}$ and $b$.  Eqs.(12) and (13) give the transition amplitude to a ghost, antighost state of negative norm.  Just as in the previous example the answer is gauge dependent.

In covariant gauges, the gauge independence of transition amplitudes between physical states is known to follow from BRST symmetry, in contrast to this the transition amplitude from physical states to ghost states depends on the gauge.  The gauge dependence in Eqs.(6), (7), (11) and (12) is a non-perturbative result arising the BRST invariance of physical wave functions.  For instance, for an infinitesimal BRST transformation $\delta$,
\begin{equation}
\delta \langle o \mid T \left(\overline{C}^{a_{1}}(x_{1}) \, A^{\mu_{2},a_{2}}(x_{2}) \, \partial_{\mu_{3}}A^{\mu_{3},a_{3}}(x_{3}) \, \partial_{\mu_{4}}A^{\mu_{4},a_{4}}(x_{4}) \right) \mid o \rangle=o
\end{equation}
from which it follows that
\begin{displaymath}
\langle o \mid T \left( \delta \overline{C}^{a_{1}}(x_{1}) \, A^{\mu_{2},a_{2}}(x_{2}) \, \partial_{\mu_{3}}A^{\mu_{3},a_{3}}(x_{3}) \, \partial_{\mu_{4}}A^{\mu_{4},a_{4}}(x_{4}) \right) \mid o \rangle
\end{displaymath}
\begin{equation}
=\langle o \mid T \left(\overline{C}^{a_{1}}(x_{1}) \, \delta A^{\mu_{2},a_{2}}(x_{2}) \, \partial_{\mu_{3}}A^{\mu_{3},a_{3}}(x_{3}) \, \partial_{\mu_{4}}A^{\mu_{4},a_{4}}(x_{4})\right) \mid o \rangle
\end{equation}
where
\begin{displaymath}
\delta \overline{C}^{a} = - {i} \, \partial^{\mu} \, A^{a}_{\mu}
\end{displaymath}
\begin{equation}
\delta A^{a}_{\mu} = \left(D_{\mu} \, C \right)^{a}
\end{equation}
are the standard BRST variations. Letting $x^{o}_{1} , x^{o}_{2} \rightarrow +\infty$ and $x^{o}_{3}, x^{o}_{4} \rightarrow - \infty$,
we use the LSZ scattering formalism.  The lhs of Eq.(15) is proportional to the gluon-gluon scattering amplitude.  The rhs is proportional to the gluon-gluon $\rightarrow$ ghost-antighost amplitude.  The l.h.s. is not zero since only three gluons are contracted with their momenta.  This shows again that the gauge dependence of what was found in Eqs.(6) and (7) is not an artifact of the weak coupling approximation.

Covariant quantization of vector fields is known to require an enlarged Hilbert space, containing states of positive, zero and negative norm\cite{mss}. Hence the time evolution of a state of positive norm will in general yield a superposition of all three states.  Electrodynamics avoids the conceptual problem of negative norm states in two ways, by imposing the Gupta-Bleuler constraint on physical states or by showing that time-like and longitudinal photon contributions to the cross-section cancel each other.

In both cases that were dealt with above the initial state has positive norm, therefore the final state will be a Hilbert space superposition of positive and zero norm states in the first case.  Even though the latter are not observable they can not be ignored completely unless they are orthogonal to physical states of positive norm.  In the second example we find that production of ghost-antighost states of negative norm is not prohibited.

\end{document}